\documentclass[12pt]{iopart}

\usepackage{iopams}  
\usepackage{graphicx} 

\def\beq{\begin{equation}}
\def\eeq{\end{equation}}
\def\bea{\begin{eqnarray}}
\def\eea{\end{eqnarray}}

\def\eq#1{{Eq.~(\ref{#1})}}
\def\fig#1{{Fig.~\ref{#1}}}
\def\ud{\underline}

\begin{document}
\title[]{Multiplicities
  in Pb-Pb central collisions at the LHC from running coupling
  evolution and RHIC data} 
\author{Javier
  L. Albacete\footnote{This research is sponsored in part by
    the U.S. Department of Energy under Grant No. DE-FG02-05ER41377.}}
\address{Department of Physics. The Ohio State University. 191
  W. Woodruff Avenue, OH-43210, Columbus, USA. E-mail: albacete@mps.ohio-state.edu}

\vskip 0.3cm
Predictions for the pseudorapidity density of charged particles produced
in Pb-Pb central collisions at $\sqrt{s_{NN}}=5.5$ TeV
presented in \cite{JLA} are summarized. Primary gluon production in
such collisions can be computed perturbatively in the framework of
$k_t$-factorization. Under the additional assumption of local parton-hadron
duality, the rapidity density of produced charged particles in nucleus-nucleus 
collisions at energy $\sqrt{s}$ and impact parameter $b$ is given by,
\cite{Kharzeev:2001gp}:
\begin{center}
\beq
\!\!\!\!\!\!\!\!\!\!\!\!\!\!\!\!\!\!\!\!\!\!\!\!\!\!\!
\frac{dN}{dy\, d^2b}=C\frac{4\pi N_c}{N_c^2-1}\int^{p_{kin}}\frac{d^2p_t}{p_t^2}\int^{p_t}\,d^2k_t\, \alpha_s(Q)\,\varphi\left(x_1,\frac{\vert\ud{k_t}+\ud{p_t}\vert}{2}\right) \varphi\left(x_2,\frac{\vert\ud{k_t}-\ud{p_t}\vert}{2}\right),
\label{ktfact}
\eeq
\end{center}
where $p_t$ and $y$ are the transverse momentum and rapidity of the
produced particle, $x_{1,2}=(p_t/\sqrt{s})\,e^{\pm y}$ and
$Q\!=\!0.5\max\left\{\vert p_t\pm k_t\vert \right\}$. The lack of
impact parameter integration in this calculation and the gluon to
charged hadron
ratio are accounted for by the constant $C$, which sets the normalization.  
The nuclear unintegrated gluon distributions (u.g.d.), $\varphi(x,k)$,
entering \eq{ktfact} are 
taken from numerical solutions of the Balitsky-Kovchegov evolution
equation including running coupling corrections, \cite{Albacete:2007yr}:
\beq
\qquad\qquad\frac{\partial N(Y,r)}{\partial
  Y}=\mathcal{R}[N(Y,r)]-\mathcal{S}[N(Y,r)]
\label{evol}
\eeq
Explicit expressions for the {\it running}, $\mathcal{R}[N]$, and {\it
  subtraction}, $\mathcal{S}[N]$, functionals in the r.h.s. of
\eq{evol} can be found in \cite{Albacete:2007yr}. The nuclear u.g.d.
are given by the Fourier transform of the dipole scattering amplitude
evolved according to \eq{evol}, $\varphi(Y,k)=\int \frac{d^2r}{2\pi
  r^2}\,e^{i\,\ud{k}\cdot\ud{r}}\,\mathcal{N}(Y,r)$, with
$Y=\ln(0.05/x)+\Delta Y_{ev}$, where $\Delta Y_{ev}$ is a free
parameter.
Large-$x$ effects have been included by replacing $\varphi(x,k)\rightarrow
\varphi(x,k)(1-x)^4$. 
The initial condition for the evolution is taken from the
McLerran-Venugopalan model \cite{McLerran:1993ni}, which is believed
to provide a good 
description of nuclear distribution functions at moderate energies:
\beq
N^{MV}(Y=0,r)=1-\exp\left\{-\frac{r^2Q_0^2}{4}\ln\left(\frac{1}{r\Lambda}+e\right)\right\}, 
\eeq
where $Q_0$ is the initial saturation scale and $\Lambda\!=\!0.2$ GeV.
In order to compare \eq{ktfact} with
experimental data it is
necessary to correct the difference between rapidity, $y$, and the
experimentally measured pseudorapidity, $\eta$. This is managed by
introducing an effective hadron mass, $m_{eff}$. The variable
transformation, $y(\eta,p_t,m_{eff})$, and its corresponding jacobian
are given by Eqs.(25-26) in 
\cite{Kharzeev:2001gp}. Corrections to the kinematics due to
the hadron mass are also considered by replacing $p_t\rightarrow
m_t=\left(p_t^2+m_{eff}^2\right)^{1/2}$ in the evaluation of
$x_{1,2}$. This replacement affects the predictions for the LHC by less than a
$5\%$, \cite{JLA}.

The results for the pseudorapidity density of charged particles in
central Au-Au collisions at
$\sqrt{s_{NN}}\!=\!130$, $200$ and $5500$ GeV are shown in \fig{pred}. A
remarkably good description 
of RHIC data is obtained with $Q_0\!=\!0.75\div 1.25$ GeV, $\Delta
Y_{ev}\!\lesssim\!3$ and $m_{eff}\!=\!0.2\div0.3$ GeV. Assuming no
difference between Au 
and Pb nuclei, the extrapolation of the fits to RHIC data yields
the following band:
$\frac{dN^{Pb-Pb}_{ch}}{d\eta}(\sqrt{s_{NN}}\!=\!5.5\,\mbox{TeV})\vert_{\eta=0}\approx 
1290\div 1480$ for central Pb-Pb collisions at the LHC. The central
value of our predictions
$\frac{dN^{Pb-Pb}_{ch}}{d\eta}(\sqrt{s_{NN}}\!=\!5.5\,\mbox{TeV})\vert_{\eta=0}\approx 
1390$ corresponds to the best fits to RHIC data.
 \begin{figure}[ht]
 \begin{center}
 \includegraphics[height=10cm]{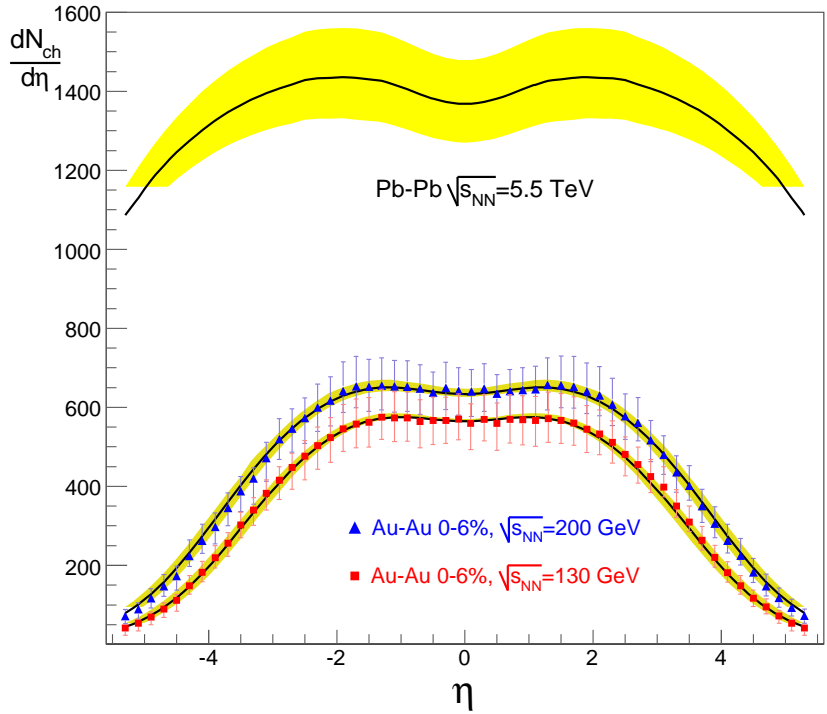}
 \vskip -0.5cm
 \caption{Multiplicity densities for Au-Au central collisions at RHIC
   (experimental data taken from \cite{Back:2004je}), and prediction
   for Pb-Pb central 
   collisions at $\sqrt{s_{NN}}\!=\!5.5$ TeV. The best fits to
   data (solid lines) are obtained with $Q_0\!=\!1$
   GeV, $\Delta Y_{ev}\!=\!1$ and $m_{eff}\!=\!0.25$ GeV. The upper limit of
   the error bands correspond  to $\Delta Y_{ev}\!=\!3$ and $Q_0\!=\!0.75$ GeV,
   and the lower limit to $\Delta Y_{ev}\!=\!0.5$ and $Q_0\!=\!1.25$ 
   GeV, with $m_{eff}\!=\!0.25$ GeV in both cases. } 
 \label{pred}
 \end{center}
 \end{figure}
\vskip -1cm

\maketitle
\end{document}